\documentclass[10pt,twocolumn,aps,prb,superscriptaddress]{revtex4-1}
\usepackage[english]{babel}
\usepackage{graphicx}
\usepackage{amsmath}
\usepackage{amssymb}
\usepackage{bm}
\usepackage[squaren,Gray]{SIunits}

\begin{document}

\title{Coulomb-driven organization and enhancement of spin-orbit fields\\ in collective spin excitations}

\author{F. Baboux}
\thanks{\texttt{baboux@insp.upmc.fr}}
\affiliation{Institut des Nanosciences de Paris, CNRS/Universit\'e Paris VI, Paris 75005, France}

\author{F. Perez}
\thanks{\texttt{perez@insp.upmc.fr}}
\affiliation{Institut des Nanosciences de Paris, CNRS/Universit\'e Paris VI, Paris 75005, France}

\author{C. A. Ullrich}
\affiliation{Department of Physics and Astronomy, University of Missouri, Columbia, Missouri 65211, USA}

\author{I. D'Amico}
\affiliation{Department of Physics, University of York, York YO10 5DD, United Kingdom}

\author{G. Karczewski}
\affiliation{Institute of Physics, Polish Academy of Sciences, Warsaw, Poland}

\author{T. Wojtowicz}
\affiliation{Institute of Physics, Polish Academy of Sciences, Warsaw, Poland}

\begin{abstract}
Spin-orbit (SO) fields in a spin-polarized electron gas are
studied by angle-resolved inelastic light scattering on a
CdMnTe quantum well. We demonstrate a striking organization and
enhancement of SO fields acting on the collective spin
excitation (spin-flip wave). While individual electronic SO
fields have a broadly distributed momentum dependence, giving
rise to D'yakonov-Perel' dephasing, the collective spin
dynamics is governed by a single \textit{collective} SO field
which is drastically enhanced due to many-body effects. The
enhancement factor is experimentally determined. These results
provide a powerful indication that these constructive phenomena
are universal to collective spin excitations of conducting
systems.
\end{abstract}

\maketitle


An electron moving with momentum $\mathbf{k}$ in an electric
field $\mathbf{E}$ experiences an effective magnetic field
$\mathbf{B}_{\rm
SO}(\mathbf{k})\propto \nolinebreak \mathbf{E}\times\mathbf{k}$ that couples
to its spin. In semiconductor nano\-structures, this spin-orbit
(SO) coupling arises through internal electric fields
\cite{WinklerBook} and opens promising ways to manipulate the
electronic spin through, e.g., current-induced spin
polarization,\cite{Kato2004b} spin Hall currents,\cite{Sih2005}
or zero-bias spin separation.\cite {Ganichev2009} However, SO
coupling generally causes also losses of spin memory: due to
the momentum dependence of SO fields, each individual spin of
an ensemble precesses with its own frequency and axis
(D'yakonov-Perel' decoherence \cite{Dyakonov1971}). In recent
years, numerous efforts have been made to overcome this
decoherence, by using structure engineering,\cite{Koralek2009}
control by gate electrodes, \cite{Studer2009,Marie2011} or
spin-echo-type techniques. \cite{Pershin2007,Kuhlen2012}

Recently, an alternative promising path was initiated
\cite{Baboux2012} by discovering that a particular collective
spin excitation, the intersubband spin plasmon of a GaAs
quantum well, is \textit{intrinsically} protected from
D'yakonov-Perel' decoherence. Indeed, Coulomb interaction
rearranges the distribution of SO fields, so that all
electronic spins precess in synchronicity about a single
\textit{collective} SO field.

In addition, this collective SO field was discovered
to be drastically enhanced with respect to the one acting on
individual electrons. Indeed, for non-collective spin
excitations such as a spin packet drifting with momentum ${\bf
q}$, the relevant SO field is that which would act on a single
electron of same momentum,
$\mathbf{B}_{\mathrm{SO}}(\mathbf{q})$.\cite{Kalevich1990,Kato2004b,Meier2007,Studer2009,Kuhlen2012}
By contrast, the SO field acting on the intersubband
spin plasmon,
$\mathbf{B}_{\mathrm{SO}}^{\mathrm{coll}}(\mathbf{q})$, is very
strongly enhanced: it was found in Ref. \onlinecite{Baboux2012}
that
$\mathbf{B}_{\mathrm{SO}}^{\mathrm{coll}}(\mathbf{q})\approx 5
\; \mathbf{B}_{\mathrm{SO}}(\mathbf{q})$.

This raises the important question whether these constructive
phenomena are bound to the peculiar nature of the intersubband
spin plasmon of a GaAs quantum well, or are fully general to
collective spin excitations of any conducting system. Here, we
evidence them in another configuration: we study the
\textit{intra}subband spin-flip wave (SFW) of a spin-polarized
electron gas confined in a diluted magnetic semiconductor
(DMS).\cite{Jusserand2003,Perez2007,AkuLeh2011} Thus, we
provide a powerful indication of the universality of the
immunity against dephasing, and giant enhancement of SO effects
at the collective level. In addition, as we shall see, the
system studied here allows for a much simpler demonstration of these effects: here, in contrast to Ref.
\onlinecite{Baboux2012}, direct observations of the SO fields
at \textit{both} the single-particle and the collective level
can be made. This allows for a fully experimental determination
of the SO enhancement factor, and shows that DMS
quantum wells are ideal systems for future study and
functionalization of collective SO effects.

\begin{figure}[tbp]
\includegraphics[width=\columnwidth]{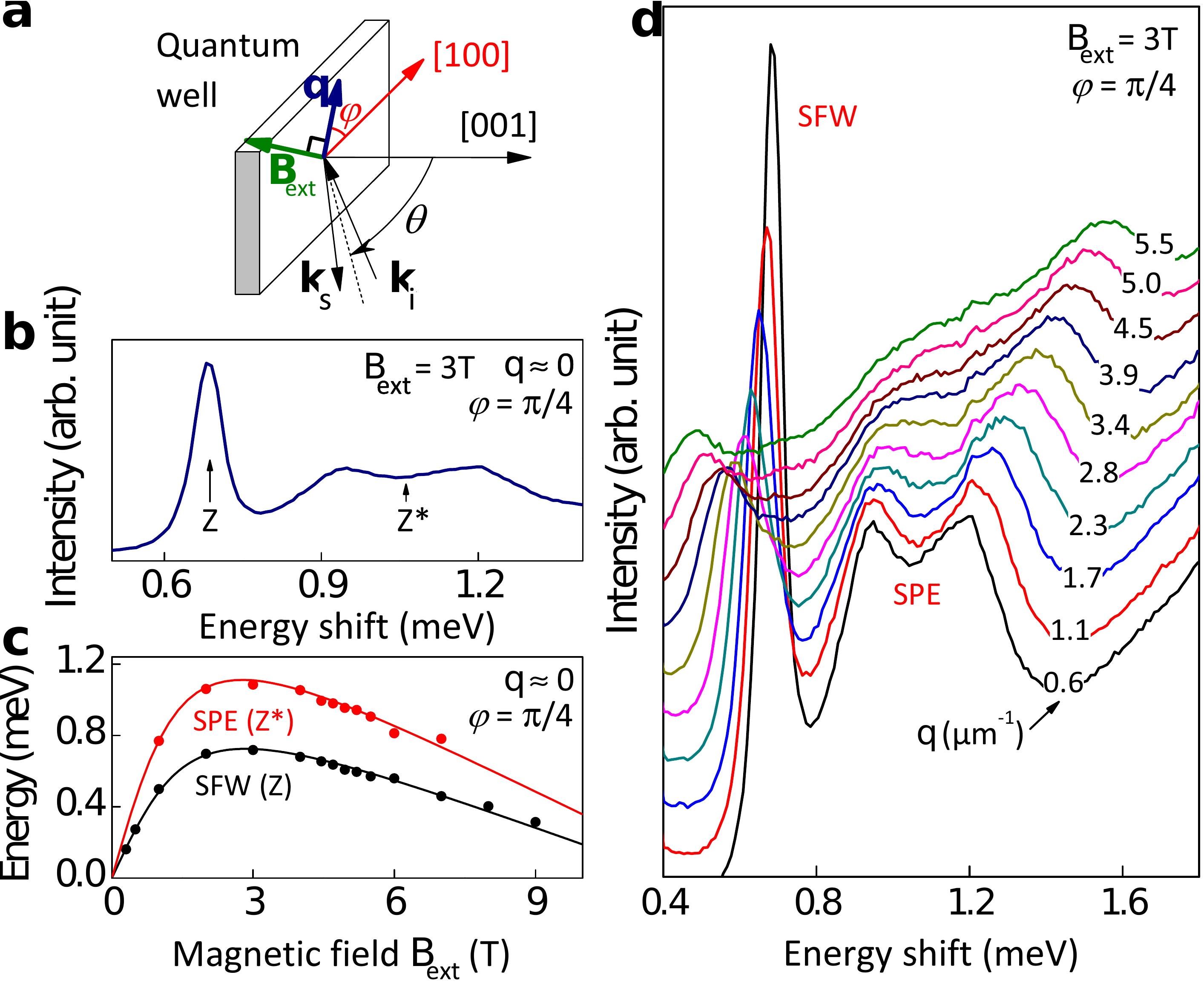}
\caption{Intrasubband
spin excitations of a CdMnTe quantum well. (a) Scattering geometry: $\mathbf{k_{i}}$ and $\mathbf{%
k_{s}}$ are the incoming and scattered light wavevectors; $\mathbf{q}$ is the transferred
momentum, of in-plane orientation $\varphi$ measured from $[100]$, and amplitude
$\vert\mathbf{q}\vert\simeq\frac{4\pi}{\lambda}\sin\theta$, where
$\lambda\simeq\unit{771}{\nano\meter}$ is the incoming wavelength.
An external magnetic field $\mathbf{B}_{\mathrm{ext}}$ is applied perpendicularly to $\mathbf{q}$
(the arrow defines $B_{\mathrm{ext}}>0$). (b) Typical inelastic light
scattering (ILS) spectrum obtained at $q \simeq 0$ in cross-polarized geometry. Two lines are
observed, corresponding to the spin-flip wave (SFW, energy $Z$) and to the spin-flip
single-particle excitations (SPE, energy $Z^{\ast }$). (c) Magnetic dispersion of $Z$
and $Z^{\ast }$. (d) ILS spectra obtained for $B_{\mathrm{ext}}=\unit{3}{\tesla}$ and a series
of transferred momenta $q$, at fixed in-plane angle $\varphi =\pi /4$.}
\label{Spectra}
\end{figure}

We carry out inelastic light scattering (ILS) measurements on
an asymmetrically modulation-doped,
$\unit{30}{\nano\metre}$-thick Cd$_{1-x}$Mn$_{x}$Te quantum
well of high mobility, grown along the $\left[001\right] $
direction by molecular beam epitaxy. The electronic density is
$n_{\mathrm{2D}}=\unit{3.5\times
10^{11}}{\rpsquare{\centi\metre}}$ and the mobility $\unit{
10^5}{\centi\squaremetre/\volt\second}$, as determined from
magneto-transport measurements. ILS is a powerful tool to
transfer a momentum ${\bf q}$ to the spin excitations of the
two-dimensional electron gas (2DEG).
\cite{Pinczuk1989,Jusserand2003} In our setup, depicted in Fig.
\ref{Spectra}(a), ${\bf q}$ can be varied both in amplitude and
in-plane orientation. A magnetic field
$\mathbf{B}_{\mathrm{ext}}$ is applied in the plane of the
well, always perpendicular to $\mathbf{q}$. $\varphi $ denotes
the angle between $\mathbf{q}$ and the $\left[ 100\right] $
crystallographic direction of the well. The incoming and
scattered light polarizations are crossed, which is the
required selection rule to address spin-flip excitations.
\cite{Pinczuk1989,Jusserand2003}

Figure \ref{Spectra}(b) shows a typical ILS spectrum obtained
for $q \simeq 0$, at superfluid helium bath temperature ($T\sim
\unit{2}{\kelvin}$). Here $B_{\mathrm{ext}}=\unit{3}{\tesla}$
and $\varphi =\pi /4$. The spectrum shows two kinds of
excitations: \cite{Jusserand2003,Perez2007} a narrow peak
corresponding to the SFW (energy $Z$) and, at higher energy, a
broader line corresponding to spin-flip single-particle
excitations\footnote{Also called Stoner excitations in magnetic metals.} (SPE, of center energy $Z^{\ast }$). The energy of
both excitations is plotted as a function of $B_{\mathrm{ext}}$
in Fig. \ref{Spectra}(c). The $q=0$ SFW involves a parallel
precession of all electron spins. Thus, its energy does not
depend on Coulomb interaction, owing to Larmor's theorem,
\cite{Jusserand2003} and follows the giant Zeeman splitting of
conduction electrons: \cite{Gaj1979}
\begin{equation}
Z(B_{\mathrm{ext}})=-\overline{x}J_{s-d}\left\langle S_{z}(B_{\mathrm{ext}})\right\rangle_{\rm th} +
g \mu _{\mathrm{B}}B_{\mathrm{ext}}. \label{Z}
\end{equation}
This is the sum of two opposite contributions (one due to the
exchange field created by the polarized Mn, and one directly
due to $B_{\mathrm{ext}}$), explaining the non-monotonic
variation of $Z$ with $B_{\mathrm{ext}}$.
$J_{s-d}=\unit{0.22}{\electronvolt}$ is the exchange integral
\cite{Gaj1979} for conduction electrons in
Cd$_{1-x}$Mn$_{x}$Te, $\overline{x}$ the effective Mn
concentration ($\overline{x}\simeq x$ for low $x$),
$\left\langle S_{z}(B_{\mathrm{ext}})\right\rangle_{\rm th}$ is
the thermally averaged spin of a single Mn$^{2+}$ ion (negative
for $B_{\mathrm{ext}}>0$), $\mu _{\mathrm{B}}$ the Bohr
magneton and  $g= -1.64$ the electronic $g$-factor. A fit of
the SFW energy to Eq. (\ref{Z}) yields $x=0.215 \, \%$ and
$T=\unit{2.6}{\kelvin}$. In contrast to the SFW, flipping the
spin of a single electron without disturbance of other spins
costs an additional Coulomb-exchange energy due to Pauli
repulsion: the center of the SPE line thus lies at a higher
energy, \cite{Perez2007} the Coulomb-renormalized Zeeman energy
$Z^{\ast }$.

We now turn to the wavevector dispersion of both excitations.
In Fig. \ref{Spectra}(d) we plot spectra obtained at fixed
$B_{\mathrm{ext}}=\unit{3}{\tesla}$ and $\varphi =\pi/4$, but
various magnitudes of the transferred momentum $\mathbf{q}$.
The energy of the SFW decreases with increasing $q$. Indeed,
spins in a $q\neq0 $ SFW mode are periodically antiparallel for
each $\pi /q$: compared to the $q = 0$ situation, this induces
a reduction in the Coulomb-exchange repulsion more and more
pronounced as $\pi /q$ is lowered, yielding a downward
dispersion. The interplay between Coulomb interaction and the
kinetic energy leads to a parabolic dispersion for the SFW
energy (to leading order in $q$):\cite{Perez2009}
\begin{equation}
E \left( q\right) =\left\vert Z-\frac{1}{ \zeta  }%
\frac{Z}{Z^{\ast }-Z}\frac{\hbar ^{2}}{2m_{b}}q^{2}\right\vert \equiv
\left\vert Z-f q^{2}\right\vert ,
\label{dispSFW}
\end{equation}
where $\zeta =(m_{b}/2\pi \hbar ^{2}n_{\mathrm{2D}})Z^{\ast }$
is the spin polarization and $m_{b}$ is the band mass.

\begin{figure}[tbp]
\includegraphics[width=\columnwidth]{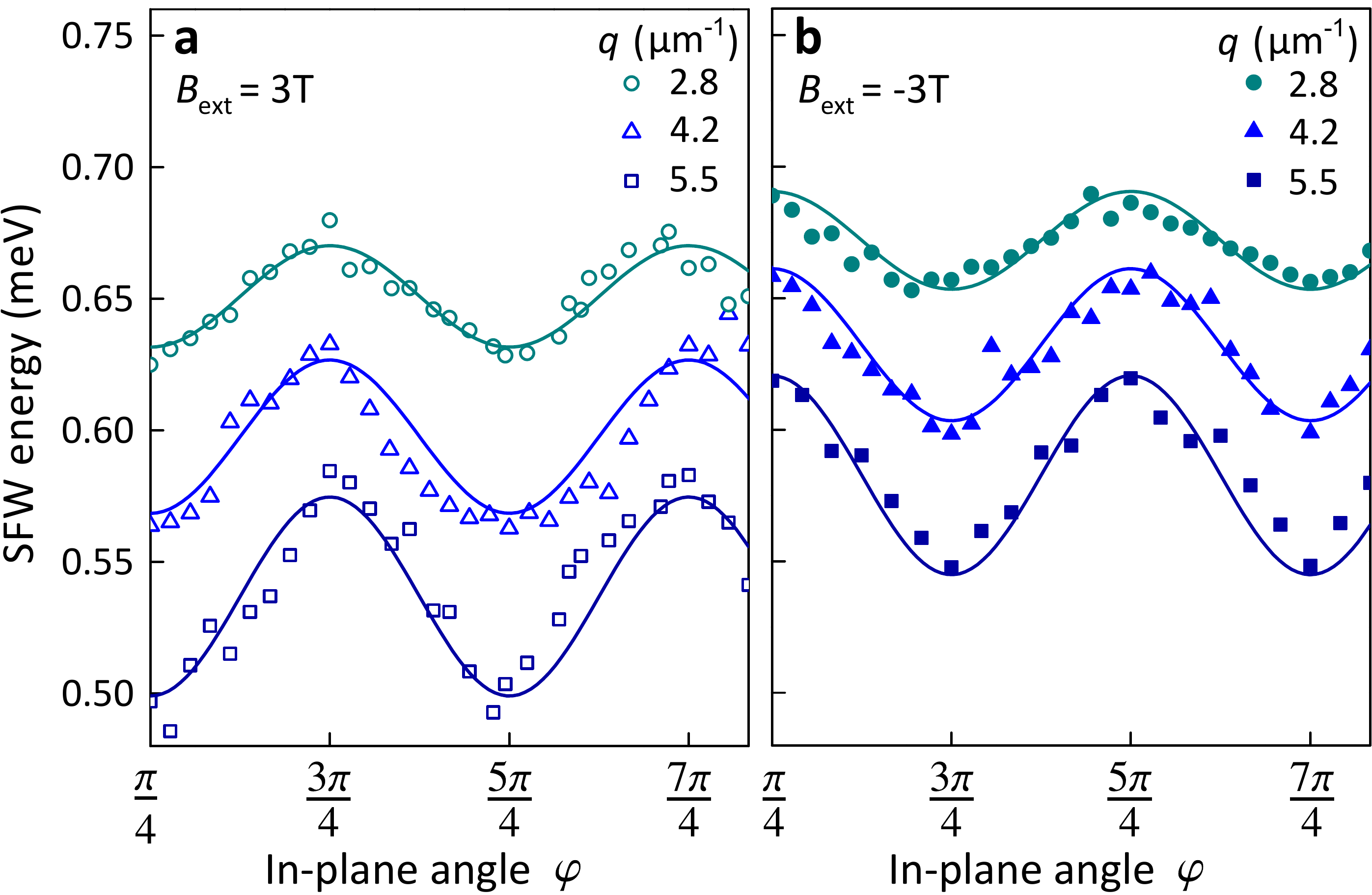}
\caption{Spin-orbit induced modulation of the spin-flip wave energy,
(a) for $B_{\mathrm{ext}}=\unit{3}{\tesla}$ and (b) $B_{\mathrm{ext}}=\unit{-3}{\tesla}$, and for three
amplitudes of the transferred momentum $q$. The modulation shows a two-fold symmetry, and its amplitude increases with $q$.
Furthermore, as compared with $B_{\mathrm{ext}}>0$, the modulation for $B_{\mathrm{ext}}<0$ is out of
phase, and lies at higher energy. The lines reproduce Eq. (\ref{En}).}
\label{Modulation}
\end{figure}

However, this expression ignores anisotropic effects stemming
from SO coupling. To investigate such effects, we plot in Fig.
\ref{Modulation} the SFW energy as a function of the in-plane
angle $\varphi$, for $B_{\mathrm{ext}}=\unit{3} {\tesla}$ (a)
and $B_{\mathrm{ext}}=\unit{-3}{\tesla}$ (b), and for three
values of the momentum $q$. Several salient features are
observed: (i) the SFW energy shows a sine-type modulation with
a twofold symmetry, analogous to the one found for
intersubband spin plasmons \cite{Baboux2012} as well as in Ref.
\onlinecite{Rice2012} for similar CdMnTe quantum wells. In
addition, (ii) the amplitude of this modulation increases with
growing $q$; (iii) the modulations obtained for
$B_{\mathrm{ext}}>0$ and $B_{\mathrm{ext}}<0$ are out of phase,
and (iv) the latter lies on average at higher energy than the
former.

To understand these features, let us consider the total
Hamiltonian of the system, containing kinetic, Coulomb,
Zeeman and SO contributions:
$\widehat{\mathcal{H}}_{\mathrm{tot}}=\widehat{\mathcal{H}}_{\mathrm{Kin}}+
\widehat{\mathcal{H}}_{\mathrm{Coul}}+\widehat{\mathcal{H}}_{\mathrm{Z}}+\widehat{\mathcal{H}}_{\mathrm{SO}}.$
The SO part reads $\widehat{\mathcal{H}}_{\mathrm{SO}}= \sum
\mathbf{B}_{\mathrm{SO}}(\mathbf{k}) \cdot
\widehat{\boldsymbol{\sigma }}/2 $, where the sum runs over all
electrons of momentum $\mathbf{k}$ and spin
$\widehat{\boldsymbol{\sigma }}$ (Pauli operators).
$\mathbf{B}_{\mathrm{SO}}(\mathbf{k})$ is an in-plane field,
whose magnitude determines the SO-induced spin
splitting at momentum $\mathbf{k}$. It arises from the
asymmetry of the confining potential (Rashba effect
\cite{Rashba1984}) and that of the crystalline cell
(Dresselhaus effect \cite{Dresselhaus1955}):
\begin{equation}
\mathbf{B}_{\mathrm{SO}}(\mathbf{k})=2\alpha \,
(k_{y},-k_{x})+2\beta \, (k_{x},-k_{y})
\label{BSOind}
\end{equation}
with $\hat{x}\parallel \left[ 100\right] $ and
$\hat{y}\parallel \left[ 010 \right] $, and with $\alpha$ and
$\beta$ the Rashba and linear Dresselhaus coupling constants,
\cite{WinklerBook} respectively.

The excitations of $\widehat{\mathcal{H}}_{\mathrm{tot}}$,
including the SFW, can in principle be calculated in linear
response theory. \cite{Ullrich2003} Instead, we will propose a
phenomenological model in line with the one validated in Ref.
\onlinecite{Baboux2012} for the intersubband spin plasmon. This
collective excitation was shown to behave as a macroscopic
quantum object of spin magnitude 1, subject to a
collective SO field proportional to the excitation momentum
$\mathbf{q}$:
\begin{equation}
\mathbf{B}_{\mathrm{SO}}^{\mathrm{coll}}(\mathbf{q})=2\widetilde{\alpha} \,
(q_{y},-q_{x})+2 \widetilde{\beta} \, (q_{x},-q_{y}) \:,
\label{BSO}
\end{equation}
where $\widetilde{\alpha }$ and $\widetilde{\beta }$ are the \textit{collective}
Rashba and Dresselhaus coupling constants, respectively.

We will assume that the SFW also behaves as a spin 1 object,
immersed in the above collective SO field, as well as the Coulomb-exchange field leading to the dispersion
of Eq. (\ref{dispSFW}). Reducing the system to the SFW
only, we are left to study the following effective Hamiltonian:
\begin{equation}
\widehat{\mathcal{H}}_{\mathrm{SFW}}=\widehat{\boldsymbol{S}}\cdot
\left[\left\vert Z-fq^{2} \right \vert\frac{\mathbf{B}_{\mathrm{ext}}}{\vert B_{\mathrm{ext}}\vert}%
+
\mathbf{B}_{\mathrm{SO}}^{\mathrm{coll}}(\mathbf{q})\right],
\label{H}
\end{equation}
where $\widehat{\boldsymbol{S}}$ is the vector of spin matrices
for a spin 1. Only the eigenstate with positive energy,
corresponding to a SFW mode with spin projection $+1$, is
addressable experimentally. \cite{Perez2009} Since the
modulation of Fig. \ref{Modulation} does not exceed $10\%$
of $Z$, we consider this eigenenergy to leading order in $\widetilde{\alpha}%
 q/Z$ and $\widetilde{\beta}  q/Z$, yielding the SFW dispersion
$E(q,\varphi)=\vert
Z-fq^{2}-2\widetilde{\alpha} q-2\widetilde{\beta}
q\sin 2\varphi \vert$. This expression reproduces the
sinusoidal modulation of the SFW energy with a two-fold
symmetry, and its increase in amplitude with $q$
[properties (i) and (ii) above].
We further note that $(Z-f q^{2})$ has the same sign as $B_{\mathrm{ext}}$; thus, if $E_{\pm}$ denotes
the SFW energy for $B_{\mathrm{ext}}\gtrless 0$,
\begin{equation}
E_{\pm }\left( q,\varphi \right) =\left\vert Z-f%
 q^{2}\right\vert \mp 2\widetilde{\alpha} q\mp 2\widetilde{\beta}%
 q\sin 2\varphi.  \label{En}
\end{equation}%
Hence $\widetilde{\beta}$ governs the
\textit{amplitude} of the SO modulation, with opposite signs
for $B_{\mathrm{ext}}>0$ and $B_{\mathrm{ext}}<0$. This
qualitatively explains property (iii) above. As for
$\widetilde{\alpha}$, it governs the \textit{energy offset}
between the two situations $B_{\mathrm{ext}}>0$ and
$B_{\mathrm{ext}}<0$, explaining property (iv).
\begin{figure}[tbp]
\includegraphics[width=\columnwidth]{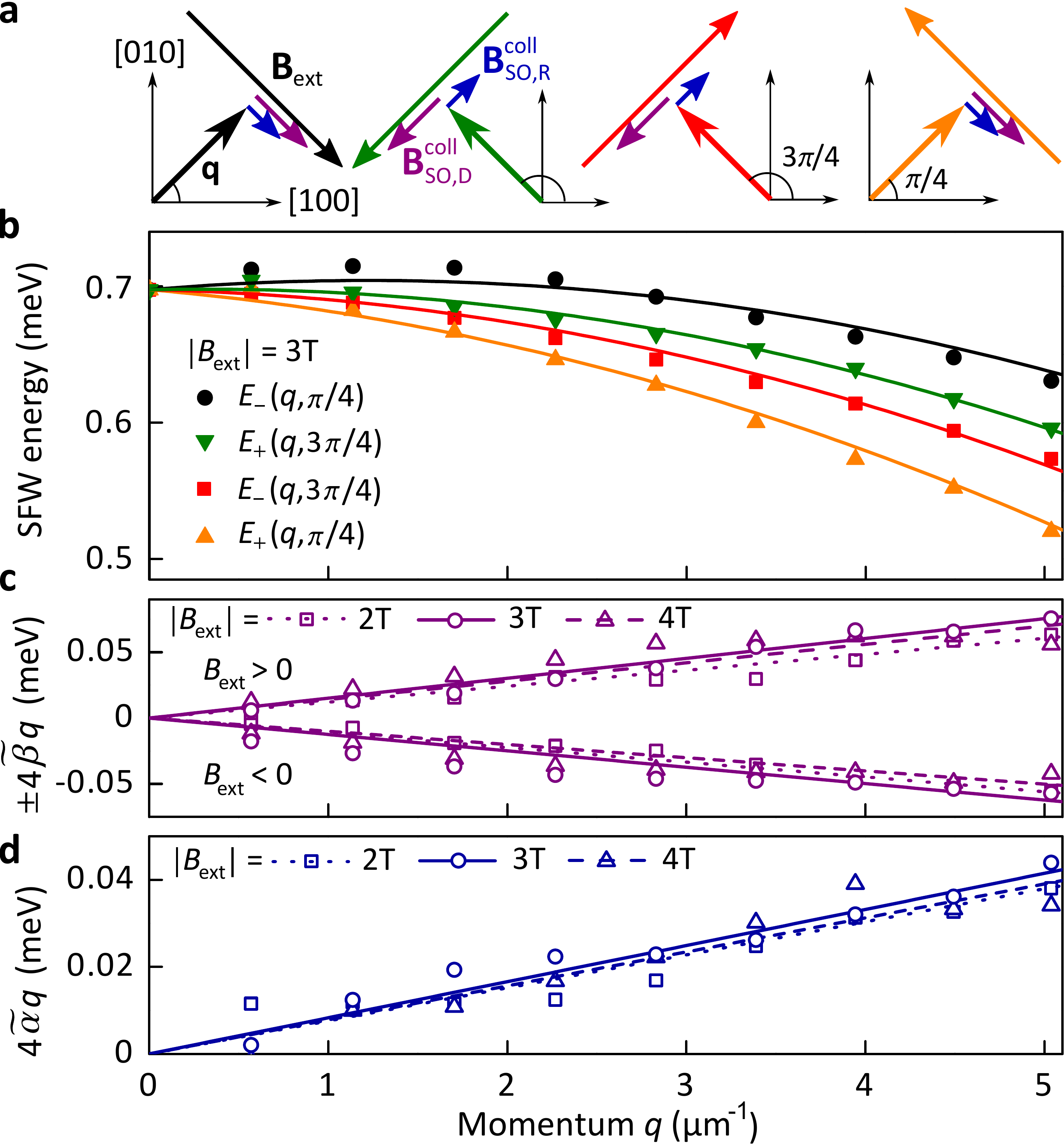}
\caption{Extraction of the collective spin-orbit coupling constants.
(a) Schematics of the four experimental configurations $(\varphi,B_{\mathrm{ext}})$
used to extract $\widetilde{\alpha}$ and $\widetilde{\beta}$. $\mathbf{B}_{\mathrm{SO,R}}^{\mathrm{coll}}$
is the Rashba component of the collective SO field (blue), and $\mathbf{B}_{\mathrm{SO,D}}^{\mathrm{coll}}$ the Dresselhaus component (purple).
(b) SFW dispersions obtained for the configurations depicted above (same color code). Lines correspond to Eq. (\ref{En}).
(c) For $\left\vert B_{\mathrm{ext}} \right\vert=2$,\,$3$ and $\unit{4}{\tesla}$, plots of $%
E_{\pm} \left( q,3\protect\pi/4
\right) -E_{\pm}\left( q,\pi /4 \right)$. Both
quantities are linear in $q$.
(d) For the same values of $\left\vert B_{\mathrm{ext}} \right\vert$, plots
of the quantity $\left\langle E_{-}\right\rangle
(q)-\left\langle E_{+}\right\rangle(q)$.}
\label{Dispersion}
\end{figure}

In the following, we will consider four specific
experimental situations that allow us to extract
$\widetilde{\alpha}$ and $\widetilde{\beta} $.
These are schematized in Fig. \ref{Dispersion}(a): for $\varphi
=\pi /4$ and $\varphi =3\pi /4$, the Rashba component of the
collective SO field
($\mathbf{B}_{\mathrm{SO,R}}^{\mathrm{coll}}$), the Dresselhaus
component ($\mathbf{B}_{\mathrm{SO,D}}^{\mathrm{coll}}$), and
the external field are collinear, yielding extrema in the SFW
energy. Figure \ref{Dispersion}(b) shows the corresponding
experimental dispersions (the same color code is used),
obtained at $\vert B_{\mathrm{ext}}\vert=\unit{3}{\tesla}$. For
$B_{\mathrm{ext}}>0$, the energy difference between $\varphi
=\pi /4$ and $\varphi =3\pi /4$, $E_{+}\left( q,3\pi /4\right)
-E_{+}\left( q,\pi /4\right)$, yields
$4\widetilde{\beta} q$ according to Eq.~(\ref{En}).
Similarly, the same difference for $B_{\mathrm{ext}}<0$ is
$-4\widetilde{\beta} q$. We plot the latter quantities
in Fig. \ref{Dispersion}(c) for previous data (circles) as well
as for additional data taken at $\left\vert B_{\mathrm{ext}}
\right\vert=2$ and $\unit{4}{\tesla}$ (squares and triangles).
A linear behavior is indeed observed, independent
of the external magnetic field within the error, in agreement with Eq.~(\ref{En}).
Averaging over all magnetic field values, we find
$\widetilde{\beta}=\unit{31.6\pm
4.5}{\milli\electronvolt\mathring{A}}$.

The collective Rashba coefficient
$\widetilde{\alpha}$ can be extracted from the energy offset
between $B_{\mathrm{ext}}>0$ and
$B_{\mathrm{ext}}<0$. Indeed, if $\left\langle E_{\pm
}\right\rangle(q)=[E_{\pm}\left( q,3\pi /4\right)
+E_{\pm }\left( q,\pi
/4\right) ]/2$ denotes the angular average of the SFW energy, the difference $\left\langle E%
_{-}\right\rangle (q)-\left\langle E%
_{+}\right\rangle (q)$ equals $4\widetilde{\alpha} q$
according to Eq.~(\ref{En}). This quantity is plotted in Fig.
\ref{Dispersion}(d) for $\left\vert B_{\mathrm{ext}}
\right\vert=2,3$ and $\unit{4}{\tesla}$. Again, very good agreement
is found with the predicted linearity, independent of the
applied magnetic field. We deduce
$\widetilde{\alpha}=\unit{19.9\pm
2.5}{\milli\electronvolt\mathring{A}}$. 

Finally, $Z$ and the
quadratic coefficient $f$ of the dispersion are determined from
the mean dispersion $[\left\langle E_{-}\right\rangle
(q)+\left\langle E_{+}\right\rangle (q)]/2$, where all SO terms
are averaged out. The consistency of the model defined by Eq.
(\ref{H}) is demonstrated in Figs. \ref{Modulation}(a)--(b) and
Fig. \ref{Dispersion}(b), where we plot the relation of Eq.
(\ref{En}) (lines) using the above extracted parameters; the
experimental SFW energy is very well reproduced.

\begin{figure}[tbp]
\includegraphics[width=\columnwidth]{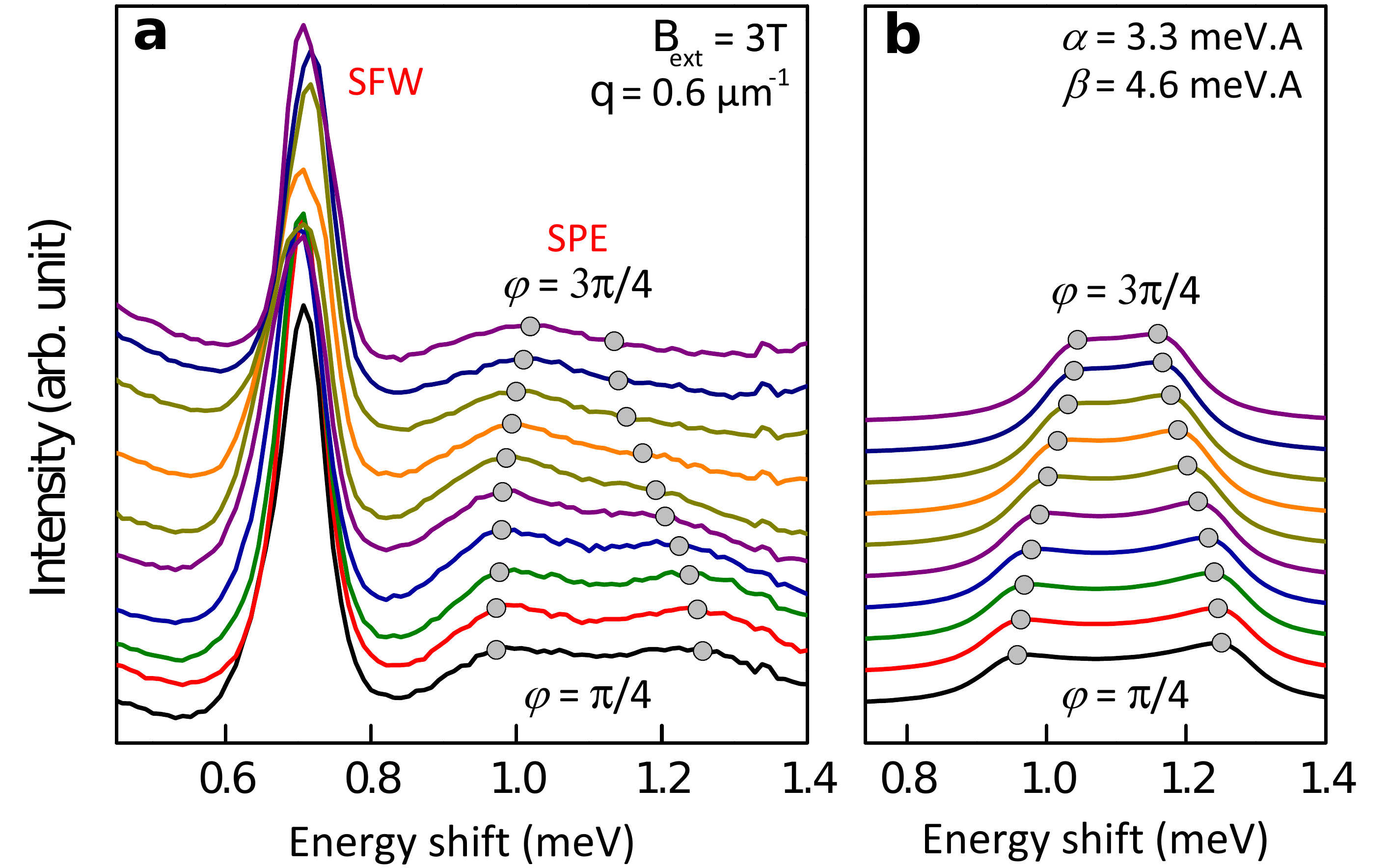}
\caption{Extraction of the individual spin-orbit coupling constants. %
(a) Experimental ILS spectra obtained at $B_{\mathrm{ext}} =\unit{3}{\tesla}$ and $q=\unit{0.6}{\
\reciprocal{\micro\metre}}$, for a series of equally spaced in-plane orientations between
$\varphi =\pi /4$ (bottom curve) and $\varphi =3\pi /4$ (top curve).  (b) With the same color code,
calculated imaginary part of the Lindhard polarizability of Eq. (\ref{Lindhard}) using $\alpha =\unit{3.3}{%
\milli\electronvolt\mathring{A}}$ and $\beta =\unit{4.6}{\milli\electronvolt \mathring{A}}$.
The grey points are guides to the eye.}
\label{SPE}
\end{figure}

The above determination of $\widetilde{\alpha}$ and
$\widetilde{\beta}$ highlights the key advantage of the
spin-polarized 2DEG to study collective SO effects: In contrast
to the intersubband spin plasmons studied in Ref.
\onlinecite{Baboux2012}, collective SO effects here directly
show up as a modulation of the SFW energy [see Eq.~(\ref{En})]. In addition, as we shall now see, SO effects at
the single-particle level can here be experimentally resolved.
Figure \ref{SPE}(a) shows spectra obtained at $B_{\mathrm{ext}}
=\unit{3}{\tesla}$ and $q=\unit{0.6}{\
\reciprocal{\micro\metre}}$, for a series of equally spaced
orientations between $\varphi =\pi /4$ (bottom curve) and
$\varphi =3\pi /4$ (top curve).
A strong in-plane modulation of the SPE line occurs, as a
result of the interplay between the external magnetic field and
the internal SO fields. The latter modulate the Fermi contour
of each spin population, with a twofold in-plane symmetry.
\cite{Badalyan2009} Van Hove-type singularities in the joint
density of states appear when both contours are locally
parallel. \cite{Jusserand1992} This gives rise to the two
shoulders observed in the SPE line.\footnote{This splitting has
to be distinguished from the splitting due to the change of
kinetic energy between the initial and final electronic states.
This quantity, of order $\hbar v_{\rm F}q$ (with $v_{\rm F}$
the Fermi velocity), is relevant at bigger $q$ as shown in Ref.
\onlinecite{AkuLeh2011}. Here in Fig. 4 $\hbar v_{\rm F}q \simeq
\unit{0.06}{\milli\electronvolt}$ is very small in comparison
to the SO-induced splitting.} The separation between
both shoulders reflects the spread of single-particle SO fields
due to their momentum dependence, in strong contrast with the
SFW, which produces a sharp line. This provides a
clear manifestation of the organization of SO fields
at the collective level.

To extract $\alpha $ and $\beta $ [see Eq. (\ref{BSOind})], we
calculate the spin-flip Lindhard polarizability
\cite{VignaleBook}
\begin{equation}
\Pi _{\downarrow \uparrow}(\mathbf{q},\omega )=\int \frac{d^{2}\mathbf{k}}{(2\pi )^{2}}\frac{%
n(\epsilon _{\mathbf{k}\downarrow })-n(\epsilon _{\mathbf{k+q}\uparrow })}{%
\epsilon _{\mathbf{k}\downarrow }-\epsilon _{\mathbf{k+q}\uparrow }+\hbar
\omega +i\eta }, \label{Lindhard}
\end{equation}%
whose imaginary part describes the shape of the spin-flip SPE
line \cite{Jusserand1992} at transferred momentum $\mathbf{q}$.
$n$ denotes the Fermi occupation function, $\epsilon
_{\mathbf{k}\sigma }$ is the energy of a single-particle state
of momentum $\mathbf{k}$ and spin $\sigma=\,\uparrow $ or
$\downarrow $, \footnote{Since $q$ is negligible in comparison
to the Fermi momentum $k_{\rm F} \sim \unit{150}{\
\reciprocal{\micro\metre}}$ and $Z^{\ast} \gg \alpha k_{\rm F}$
and $\beta k_{\rm F}$, the change of spin orientation between
the initial and final electronic states can be neglected.} and
$\eta $ accounts for the finite lifetime $\hbar /\eta $ of
quasi-electrons due to scattering off disorder and other
electrons. In the approximation of strong external field used
above [see Eq.~(\ref{En})], $\epsilon _{\mathbf{k}\sigma
}\simeq \frac{\hbar ^{2}k^{2}}{2m_{b}}+\sigma
\,\mathrm{sgn}(B_{\mathrm{ext}}) \big(Z^{\ast }/2-\big[\alpha
\cos \left( \varphi -\varphi_{\mathbf{k}}\right) +\beta \sin
\left( \varphi +\varphi_{\mathbf{k}}\right) \big]k\big)$, where
$\mathrm{sgn}(B_{\mathrm{ext}})$ is the sign of
$B_{\mathrm{ext}}$, $\sigma=\pm 1$ and $\varphi_{\mathbf{k}}$ is the angle between
$\mathbf{k}$ and the $\left[ 100\right] $ direction of the
well.

Figure \ref{SPE}(b) shows the calculated imaginary part of $\Pi
_{\downarrow \uparrow}$ for the experimental parameters of Fig.
\ref{SPE}(a) (same color code). The main experimental trend is
well reproduced  by using $\alpha =\unit{3.3}{
\milli\electronvolt\mathring{A}}$, $\beta
=\unit{4.6}{\milli\electronvolt \mathring{A}}$ and $\eta
=\unit{0.05}{\milli\electronvolt}$. Note that the fitted
disorder parameter $\eta$ (which affects the softening of the
line shape, but not the magnitude of the splitting) is at least
three times lower than for previously investigated samples,
\cite{Perez2007,Gomez2010,AkuLeh2011} confirming a very high sample
quality. Indeed, Fig. \ref{SPE}(a) is to our knowledge the
first ILS observation of a SO splitting in the SPE line of a
DMS quantum well.

The measured $\alpha$ and $\beta $ can be compared to
theoretical estimates. The Rashba coefficient can be calculated
from $\alpha =r_{41}^{6c6c}e\langle E_{z}\rangle $, with $e$
the electronic charge and $\langle E_{z}\rangle $ the average
electric field along the growth axis. Assuming that the
electrons experience the delta-doping layer as an infinite
sheet of positive charge, and using
$r_{41}^{6c6c}=\unit{6.93}{\mathring{A}^{2}}$ calculated by
$\mathbf{k\cdot p}$ perturbation theory \cite{WinklerBook} for CdTe, we obtain
$\alpha_{\mathrm{kp}}=\unit{2.2}{\milli\electronvolt\,\angstrom}$.
The Dresselhaus coefficient reads $\beta =\gamma \langle
k_{z}^{2}\rangle $.
Using $\gamma =\unit{43.9}{\electronvolt\mathring{A}^{3}}$ from $\mathbf{%
k\cdot p}$ theory \cite{WinklerBook} and estimating
$\langle k_{z}^{2}\rangle$ for a square well, we find
$\beta_{\mathrm{kp}}=\unit{4.7}{\milli\electronvolt\,\angstrom}$. Hence, the
above experimental determination of $\alpha$ and $\beta$ is very
well supported by these simple estimates.

We are now in a position to make a direct comparison between
the magnitude of SO effects at the individual and at the
collective level. We find $\widetilde{\alpha }\sim
6\,\alpha$ and $\widetilde{\beta}\sim 7\,\beta$, so that
$\mathbf{B}_{\mathrm{SO}}^{\mathrm{coll}}(\mathbf{q} )\simeq
6.5\,\mathbf{B}_{\mathrm{SO}}(\mathbf{q})$:
The interplay of
Coulomb and SO interactions produces a striking boost of the
Rashba and Dresselhaus effects at the collective level, while
preserving the balance between both. This organization and enhancement arises mainly from
Coulomb-exchange interaction, which naturally tends to align
spins: it gives rise to an additional $\mathbf{k}$-dependent
magnetic field \cite{Ullrich2003} that exactly compensates the
$\mathbf{k}$-dependence of SO fields, together with enhancing
their common component aligned with by
$\mathbf{B}_{\mathrm{SO}}(\mathbf{q})$.

In conclusion, using a test-bed spin-polarized 2DEG, we have
carried out direct optical measurements of SO fields at the individual and at the
collective level. The broad, split line of single-particle
excitations reflects the spread of individual SO fields due to
their momentum dependence. In strong contrast, the SFW remains
a sharp line reflecting the precession of a macroscopic spin in
a single collective SO field proportional to the excitation
momentum ${\bf q}$. Due to many-body effects, this field is
drastically enhanced with respect to the one acting on
individual electrons. Together with the findings of Ref.
\onlinecite{Baboux2012}, these results provide a powerful
indication that the observed phenomena are universal to
collective spin excitations in conducting systems. This
remarkable behavior provides a strong incentive for studying
these effects in other helical liquids \cite{Raghu2010} such as
in topological insulators, where SO coupling is very large.
\cite{King2011}

We thank M.~Bernard and S.~Majrab for technical support and
B.~Jusserand for fruitful discussion. F.B. is supported by a
Fondation CFM-JP Aguilar grant. F.P. acknowledges funding from
C'NANO IDF and ANR. C.A.U. is supported by DOE Grant
DE-FG02-05ER46213. I.D'A. acknowledges support from EPSRC Grant
EP/F016719/1 and I.D'A. and F.P. acknowledge support from Royal
Society Grant IJP 2008/R1 JP0870232. The research in Poland was
partially supported by the European Union within European
Regional Development Fund, through grant Innovative Economy
(POIG.01.01.02-00-008/08).


%

\end{document}